\documentclass[reprint,aps,prl,10pt,twocolumn]{revtex4-1}

\usepackage{amsmath}
\usepackage{amsfonts}
\usepackage{amssymb}
\usepackage{mathtools}
\usepackage{braket}
\usepackage{bm}
\usepackage{graphicx}
\usepackage{sansmath}

\usepackage{hyperref}

\hypersetup{
	colorlinks=true,
	linkcolor=blue,
	citecolor=blue,
	filecolor=magenta,
	urlcolor=cyan
}

\let\mod\relax
\DeclareMathOperator{\mod}{\, \text{mod}\,}

\graphicspath{{../}}

\newcommand{\papertitle}{Topology of one dimensional quantum systems out of equilibrium}
\newcommand{\tcm}{T.C.M. Group, Cavendish Laboratory, University of Cambridge, JJ Thomson Avenue, Cambridge, CB3 0HE, U.K.}

\allowdisplaybreaks[1]

\DeclareSymbolFont{sfletters}{OML}{cmbrm}{m}{it}

\newcommand{\matr}[1]{\mathsf{#1}}
\DeclareMathSymbol{\matrrho}{\mathord}{sfletters}{"1A}

\begin{document}
	
	\title{\papertitle}
	\author{Max McGinley}
	\affiliation{\tcm}
	\author{Nigel R. Cooper}
	\affiliation{\tcm}
	
	\date{\today}
	
	\begin{abstract}
		We study the topological properties of one-dimensional systems undergoing unitary time evolution. 
		We show that symmetries possessed both by the initial wavefunction and by the Hamiltonian at all times may not be present in the time-dependent wavefunction -- a phenomenon which we dub ``dynamically-induced symmetry breaking". 
		This leads to the possibility of a time-varying bulk index after quenching within non-interacting gapped topological phases. The consequences are observable experimentally through particle transport measurements. With reference to the entanglement spectrum, we explain how the topology of the wavefunction can change out of equilibrium, both for non-interacting fermions and for symmetry-protected topological phases protected by antiunitary symmetries. 
	\end{abstract}
	
	\maketitle
	
	\bibliographystyle{apsrev4-1.bst}
	
	In the past few decades, numerous examples of gapped quantum many-particle systems with topologically non-trivial ground states  have been discovered \cite{Hasan2010, Qi2011}. 
	Despite the lack of a local order parameter, these states cannot be smoothly connected to their topologically trivial counterparts without closing the bulk energy gap and removing their characteristic gapless edge modes.	
	
	Central to the modern understanding of these phases is the importance of symmetry constraints on the Hamiltonian, through which a rich `periodic table' of non-interacting fermionic topological phases emerges \cite{Qi2008,Schnyder2008,Kitaev2009,Ryu2010}. Such systems can be characterized by bulk indices which capture global features of the Bloch bands, generalizing the Chern number for two-dimensional systems \cite{Thouless1982}. These indices are topological invariants: they are unchanged under symmetry-respecting deformations of the Hamiltonian, provided the gap does not close.  
	More general symmetry-protected topological (SPT) phases are also known to exist beyond free fermions \cite{Xie2013}.
	
	More recently, the topological properties of quantum states far from equilibrium have been examined \cite{Patel2013,Kells2014,Hauke2014,Mazza2014,Calvanese2016,Wang2017,Foster2013,Sacramento2014,DAlessio2015,Caio2015,Yang2018,Gong2017,Liu2018}, motivated by possibilities to study coherent dynamics in cold atom experiments \cite{Flaschner2018,Tarnowski2017,Cooper2018}. The Chern number after a quantum quench has been shown to be constant in time \cite{Foster2013,Sacramento2014,DAlessio2015,Caio2015}, a result that  
	has often been assumed to be a universal feature of all bulk invariants in non-interacting fermionic systems \cite{Yang2018,Liu2018}. However, existing studies leave open the role of symmetry in the post-quench state.
	
	In this paper we address the effects of symmetries on the topology of one dimensional (1D) quantum systems that are out of equilibrium. We show that the bulk index of the time-evolved wavefunction \textit{can} vary in time. Surprisingly, this can occur {even} when the Hamiltonian retains the required symmetries at all times and remains within the same phase. This behaviour stems from a phenomenon which we call ``dynamically-induced symmetry breaking": after a quantum quench, the symmetries of the time-dependent state do not necessarily match those of the governing Hamiltonian. We determine the dynamical behaviour of the bulk index in all symmetry classes for non-interacting fermions in 1D, and show that the predicted dynamics of the bulk index can be directly measured in experiment. We also describe how the bulk index relates to the topology of the wavefunction out of equilibrium, using the entanglement spectrum \cite{Li2008}.

	We conclude by explaining the relevance of dynamically-induced symmetry breaking to interacting SPT phases, and numerically demonstrate the consequences for the entanglement spectrum of time-reversal protected Haldane phases. Our work highlights the difference between static and dynamic protection of topological phases in general: whilst the topological properties of a ground state may be robust against time-independent symmetry-respecting perturbations, the same is not necessarily true of time-dependent symmetry-respecting perturbations.

	\emph{Symmetry under dynamics.---} At equilibrium, non-interacting fermionic topological insulators are classified into ten symmetry classes according to the presence of the `generic' symmetries of time-reversal (TRS), particle-hole (PHS), and chiral (or sublattice) symmetry \cite{Altland1997,Schnyder2008}. Note that in superconducting systems, PHS is not a physical symmetry, but represents a redundancy in the Bogoliubov-de Gennes equations \cite{Bernevig2013}. Each of these symmetries imposes a constraint on the matrix $H_{ij}$ that defines the Hamiltonian $\hat{\mathcal{H}}$ via $\hat{\mathcal{H}} = \hat{\psi}_i^\dagger H_{ij} \hat{\psi}_j$ where $\hat{\psi}_j^{\dag}$ creates a fermion in a state $j$. These are \cite{Chiu2016}
	\vspace{-2pt}\begin{subequations}
		\label{eqFirstQuantizedSymmetries}
		\begin{align}
		\matr{T} \matr{H}^* \matr{T}^\dagger &= \matr{H} & \text{TRS} 
		\label{eqFirstQuantizedSymmetries:T}
		\\
		\matr{C} \matr{H}^* \matr{C}^\dagger &= -\matr{H} & \text{PHS}
		\label{eqFirstQuantizedSymmetries:C}
		\\
		\matr{S} \matr{H} \matr{S}^\dagger &= -\matr{H} & \text{Chiral}
		\label{eqFirstQuantizedSymmetries:S}
		\end{align}
	\end{subequations}
	\vspace{-1pt}\noindent where $\matr{T}, \matr{C}, \matr{S}$ are unitary matrices that satisfy $\matr{T}^* \matr{T} = \pm \matr{1}$, $\matr{C}^* \matr{C} = \pm \matr{1}$, and $\matr{S}^* \matr{S} = \matr{1}$. 
	
	For systems with a unique ground state, the symmetries \eqref{eqFirstQuantizedSymmetries} of the Hamiltonian are inherited by the ground state wavefunction $\ket{\Psi}$, and therefore by the single-particle density matrix $\rho_{ij} = \braket{\Psi|\hat{\psi}_i^\dagger \hat{\psi}^{\phantom{\dag}}_j|\Psi}$, which itself fully characterizes the state of non-interacting fermions. One finds \cite{Ryu2010}
	{
		\allowdisplaybreaks[0]
		\begin{subequations}
			\label{eqEquibSymmetryRho}
			\begin{align}
			\matr{T} \matrrho^* \matr{T}^\dagger &= \matrrho & \text{TRS} 	\label{eqEquibSymmetryRho:T}
			\\
			\matr{C} \matrrho^* \matr{C}^\dagger &= \matr{1} - \matrrho & \text{PHS}	\label{eqEquibSymmetryRho:C}
			\\
			\matr{S} \matrrho \matr{S}^\dagger &= \matr{1} - \matrrho & \text{Chiral}
			\label{eqEquibSymmetryRho:S}
			\end{align}
		\end{subequations}
	}
	This characterization of the symmetry properties of the state \eqref{eqEquibSymmetryRho} admits a natural generalization out of equilibrium. We consider non-equilibrium states arising from a very general quench protocol: the system is prepared in the ground state of an initial Hamiltonian $\matr{H}^\text{i}$ at time $t=0$ and then evolves under some other Hamiltonian $\matr{H}^\text{f}(t)$, which may itself vary in time in an arbitrary manner. The single particle density matrix evolves as $\matrrho(t) = \matr{U}(t) \matrrho(0) \matr{U}(t)^\dagger$ under the time evolution matrix $\matr{U}(t) = \mathcal{T} \exp [-i\int_0^t dt'\, \matr{H}^\text{f}(t')]$ ($\mathcal{T}$ denotes time-ordering). By replacing $\matrrho$ with $\matrrho(t)$ in  \eqref{eqEquibSymmetryRho}, we can determine the symmetries of the state at time $t$.
	
	We find two general mechanisms by which the symmetries of the initial state can be broken for $t>0$.
	
	{\it Explicit Symmetry Breaking.---} If a symmetry of the Hamiltonian changes between $\matr{H}^\text{i}$ and
	$\matr{H}^\text{f}(t)$, this symmetry will not appear in the state at $t > 0$~\footnote{There do exist
		fine-tuned scenarios where symmetry is preserved even when
		$H^\text{f}(t)$ breaks the symmetry \cite{Mazza2014}, but we consider
		only general cases here.}.  This applies in simple situations where a generic symmetry of the Hamiltonian is lost, e.g.\@ if $\matr{H}^{\rm i}$ has chiral (sublattice) symmetry but
	$\matr{H}^{\rm f}(t)$ does not.  However, it also 
	applies in situations where a generic symmetry is preserved, but the matrix ($\matr{T}, \matr{C}$ or $\matr{S}$) that realizes the symmetry changes. For example even if  chiral (sublattice) symmetry is preserved,  the sets of sites that constitute the two sublattices could differ between $\matr{H}^\text{i}$ and $\matr{H}^\text{f}(t)$.
	(We provide other
	examples in the Supplemental Material \cite{SI}.)
	
	{\it Dynamically-Induced Symmetry Breaking.---} Even if there is no change in symmetry of the Hamiltonian -- i.e.\@ initial and final Hamiltonians have the same symmetries,  realized by the same unitary matrices --  we find that there can be a change in the symmetry of the state purely due to unitary dynamics. In this case the density matrix satisfies
	\begin{subequations}
		\label{eqNonEquibSymmetryRho}
		\begin{align}
		\matr{T} \matrrho(t)^* \matr{T}^\dagger &= \matrrho(-t) & \text{TRS} 
		\label{eqNonEquibSymmetryRho:T} \\
		\matr{C} \matrrho(t)^* \matr{C}^\dagger &= \matr{1} - \matrrho(t) & \text{PHS}	
		\label{eqNonEquibSymmetryRho:C}  \\
		\matr{S} \matrrho(t) \matr{S}^\dagger &= \matr{1} - \matrrho(-t) & \text{Chiral}
		\label{eqNonEquibSymmetryRho:S}
		\end{align}
	\end{subequations}
	\noindent 
	where we have used $\matrrho(-t)$ to denote a fictitious system time-evolved by a time $+t$ under the Hamiltonian $-\matr{H}^\text{f}(t)$. Because in general $\matrrho(-t) \neq \matrrho(t)$, we infer that, surprisingly, TRS and chiral symmetries of the state are \textit{not} preserved under dynamics, because (\ref{eqNonEquibSymmetryRho:T},\ref{eqNonEquibSymmetryRho:S}) are not equivalent to the symmetry conditions (\ref{eqEquibSymmetryRho:T},\ref{eqEquibSymmetryRho:S}). On the other hand, the time-dependent PHS condition \eqref{eqNonEquibSymmetryRho:C} \textit{is} equivalent to the equilibrium case \eqref{eqEquibSymmetryRho:C}, so PHS is the one generic symmetry that is retained at all times.

	\begin{table}
		\begin{ruledtabular}
			\begin{tabular}{l@{\hspace{0.15in}}ccc@{\hspace{0.15in}}ccc}
				Class & $\matr{T}$ & $\matr{C}$ & $\matr{S}$ & $\text{CS}_1(t=0)$ & $\text{CS}_1(t)\, \text{mod}\, 1$ & Class./Ent. \\ \colrule
				AIII & 0 & 0 & 1 & $\mathbb{Z}/2^*$ & Varies $[0,1)$ & 0 \\
				BDI & $+$ & $+$ & 1 & $\mathbb{Z}/2^*$ & Const.\@ $\{0, 1/2\}$ & $\mathbb{Z}_2$ \\
				D & 0 & $+$ & 0 & $\mathbb{Z}/2 \, \text{mod}\, 1$ & Const.\@ $\{0, 1/2\}$ & $\mathbb{Z}_2$ \\
				DIII & $-$ & $+$ & 1 & $\mathbb{Z} \, \text{mod}\, 2^*$ & Const.\@ $0$ & 0 \\
				CII & $-$ & $-$ & 1 & $\mathbb{Z}^*$ & Const.\@ $0$ & 0 \\
			\end{tabular}
		\end{ruledtabular}
		\caption{Topological characterizations of 1D insulators in and out of equilibrium. The five non-trivial classes in 1D are defined by the presence of TRS, PHS and chiral symmetries ($\matr{T}$, $\matr{C}$, $\matr{S}$) according to Eq.\@ \eqref{eqFirstQuantizedSymmetries}, and their topologically distinct values of $\text{CS}_1$ in equilibrium are given. Asterisks denote cases for which $\text{CS}_1$ must be evaluated in a gauge specified by the TRS or chiral symmetries. After time evolving under a Hamiltonian in the same symmetry class, the fractional part of $\text{CS}_1(t)$ either varies in time, or stays fixed to its initial value. The possible values of $\text{CS}_1(t) \mod 1$ are given, which determine the topological classification (Class.) out of equilibrium. Non-trivial wavefunctions within this classification will also have degenerate entanglement spectra (Ent.).}
		\label{tabClassifciation}
	\end{table}

	In the following we will focus on quantum quenches without explicit symmetry breaking. 
	
	\emph{Dynamics of the bulk index.---}	At equilibrium, the bulk index which characterizes topology in 1D is the Chern-Simons (CS) invariant \cite{Ryu2010}, or equivalently the Zak phase $\alpha_\text{Z}$ \cite{Zak1989}
	\begin{align}
	\text{CS}_1 \equiv \frac{\alpha_\text{Z}}{2\pi} \coloneqq \frac{i}{2\pi} \int_{\text{BZ}} dk\, \braket{u^\alpha_k|\partial_k u^\alpha_k},
	\label{eqChernSimons}
	\end{align}
	expressed in terms of the ground state Bloch functions $\ket{u^\alpha_k}$ for occupied bands $\alpha$ (a sum over $\alpha$ for all occupied bands is to be understood). The functions are 
	assumed to vary smoothly with wavevector $k$ and are chosen
	to be periodic in the Brillouin zone (BZ).
	
	The CS invariant is only defined modulo 1, since gauge transformations of the occupied Bloch states can change $\text{CS}_1$ by an integer. However, in the presence of TRS and/or chiral symmetry, the integer part can be given physical meaning through the use of certain symmetry-related gauge choices \cite{Ryu2010}. Under such gauges, all equilibrium topological invariants in 1D can be deduced from quantized (integer or half-integer) values of the CS invariant. These quantized values, and hence the topological classification, arise only when particular symmetry combinations are imposed. The five non-trivial classes are listed in Table \ref{tabClassifciation} with their topological classifications under CS$_1(t=0)$. We consider the effects of dynamically-induced symmetry breaking on CS$_1$ in these five classes.

	All states which possess PHS (classes BDI, D, DIII, and CII) must have a CS invariant quantized to 0 or $1/2$ up to the addition of an integer \cite{Budich2013}. As we have shown, PHS is preserved under time evolution, and so the time-dependent $\text{CS}_1(t)$ must also be quantized for $t > 0$. Moreover, assuming that all Hamiltonians are smooth in $k$-space, one can define a continuous PHS-preserving interpolation between the initial and final states parametrized by the time $t$, under which the fractional part of $\text{CS}_1(t)$ cannot change. The fractional part of $\text{CS}_1(t)$ is therefore constant when PHS is present.

	States which do not possess PHS can have a CS invariant quantized to half-integer values if there is a chiral symmetry (class AIII).
	We have argued above that chiral symmetry will in general undergo dynamically-induced symmetry breaking. Thus, for $t > 0$ the CS invariant need no longer be quantized, and one expects $\text{CS}_1(t)$ to vary in time. 
	This leads to the surprising finding that even when the initial and final Hamiltonians satisfy the same (chiral) symmetry at all times the bulk index becomes time-dependent. 	
	
	\emph{Relation to physical observables.---} 
	Remarkably, the dynamics of the bulk index has directly observable consequences even far from equilibrium. (This contrasts with the Chern index for which the
	relationship with the Hall conductance does not hold out of equilibrium \cite{Caio2016,Unal2016}.) Specifically, the identification of $\text{CS}_1$ with the bulk polarization of the system (i.e.\@ the centres of Wannier states) \cite{King1993} still holds beyond the adiabatic limit. To show this, we calculate the mean current  
	\begin{align}
	\braket{j(t)} &= \frac{1}{2\pi}\int_{\text{BZ}} \!\!\! dk\, \braket{u_k^\alpha(t)|\partial_k \hat{H}^\text{f}_k(t)| u_k^\alpha(t)} \nonumber\\
	&= \frac{1}{2\pi}\int_{\text{BZ}} \!\!\! dk\, \left\{ \partial_k \left[\braket{u_k^\alpha(t)| \hat{H}^\text{f}_k(t)| u_k^\alpha(t)}\right] \right. \nonumber\\&\left. - \braket{u_k^\alpha(t)|\hat{H}^\text{f}_k(t)|\partial_k u_k^\alpha(t)} - \braket{\partial_k u_k^\alpha(t)|\hat{H}^\text{f}_k(t)|u_k^\alpha(t)} \right\} \nonumber\\
	&= \frac{i}{2\pi}\int_{\text{BZ}} \!\!\! dk\, \left[ \braket{\partial_t u_k^\alpha(t)| \partial_k u_k^\alpha(t)} + \braket{u_k^\alpha(t)|\partial_t \partial_k u_k^\alpha(t)}\right] \nonumber\\
	&= \frac{d}{dt}\text{CS}_1
	\label{eqCurrent}
	\end{align}
	\noindent We have integrated by parts, and used the periodicity of $|u^\alpha_k\rangle$ in the BZ.  Thus, the time variation of $\text{CS}_1(t)$ is reflected in the post-quench current and bulk polarization, which can be measured in experiment. Note that no assumption of any form of adiabaticity is required.

	\begin{figure}
		\includegraphics[width=246pt]{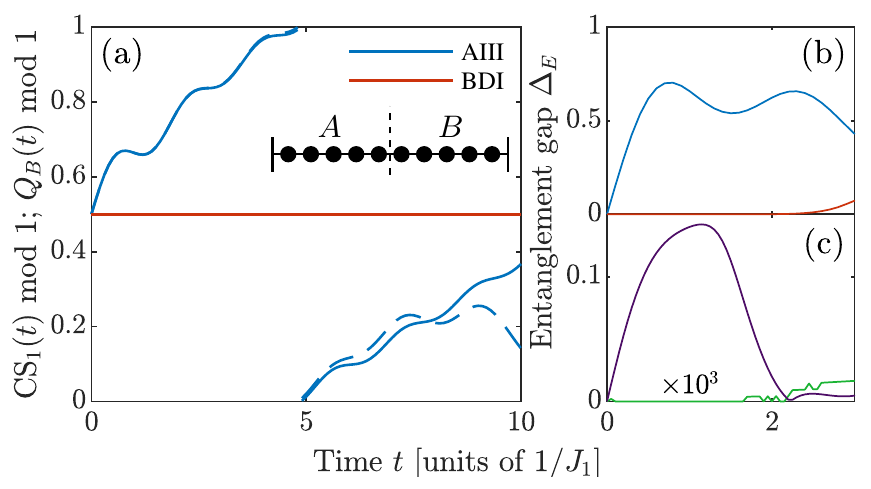}
		\caption{Panel (a): Time-dependent CS invariant of a hopping model of spinless fermions, calculated as a bulk integral in $k$-space (solid lines), compared with the polarization $Q_B(t)$ of a 24-site open boundary system with the same parameters (dashed lines). $Q_B(t)$ is calculated as the expected particle number within the righthand half, subsystem $B$ of the inset. The red lines are for a BDI system and the blue lines are for an AIII system. The parameters for the quenches are $(J_1,J_2) = (0.3,e^{i\alpha}) \rightarrow (0.8\, e^{i\alpha},1)$ with $B_{1,2} = 0.05$ throughout; $\alpha = 0$ for class BDI and $\alpha = 0.4$ for class AIII. The observables in the finite sample match the dynamics of the bulk invariants even out of equilibrium, until correlations traverse the whole system at which point the discrete nature of $k$-space invalidates Eq.\@ \eqref{eqCurrent}. Panel (b): dynamics of the entanglement gap $\Delta_E$ for the same systems as above with the entanglement cut between $A$ and $B$ [inset of (a)]. In the BDI case, the entanglement gap remains close to zero until correlations span the system size, whereas the AIII system immediately becomes gapped. Panel (c): dynamics of the entanglement gap for a spin-1 chain initialized in a Haldane phase, possessing TRS only (purple line), and both TRS and dihedral symmetry (green line, scaled by $10^3$). When the Haldane phase is supported by TRS only, the entanglement energies become gapped for $t > 0$.}
		\label{figCurrent}
	\end{figure}
	
	We have numerically verified that this relationship between the CS invariant and local current holds, even within the bulk of a finite system. We consider spinless fermions, represented by operators  $\hat{\psi}^{(\dagger)}_j$ acting on the sites labelled by $j$, with a hopping Hamiltonian $\hat{\mathcal{H}} = -\sum_{j} ( J_1\hat{\psi}_{2j+1}^\dagger \hat{\psi}_{2j}^{\phantom{\dagger}} + J_2\hat{\psi}_{2j+2}^\dagger\hat{\psi}_{2j+1}^{\phantom{\dagger}}  + B_1 \hat{\psi}_{2j+3}^\dagger\hat{\psi}_{2j}^{\phantom{\dagger}} + B_2 \hat{\psi}_{2j+4}^\dagger \hat{\psi}^{\phantom{\dagger}}_{2j+1} + \text{h.c.})$. In general, the model possesses only a chiral sublattice symmetry (class AIII), but if all hopping amplitudes are real, TRS and PHS are also present (class BDI). Figure \ref{figCurrent}(a) shows the time variation of the CS invariants for AIII and BDI systems, calculated as bulk integrals. This is compared to the bulk polarization $Q_B(t)$ in a finite system with the same hopping amplitudes, calculated as the particle number in the right subsystem $B$ (see inset). The gauge-invariant $Q_B(t)$ equals $\text{CS}_1(t)$ up to an integer, until correlations span the whole system. Thus in 1D, the change in the CS invariant is directly measurable as particle accumulation.
	
	Note that the time-variation of $\text{CS}_1$ can be seen in even simpler models such as the SSH model \cite{Su1979} with complex hopping amplitudes, i.e.\@ our model with $B_1 = B_2 = 0$ (which is in class BDI). If the phases of either $J_{1,2}$ change across the quench, then TRS and PHS undergo explicit symmetry breaking, and one finds the same behaviour as expected for an AIII quench: the dynamically-induced breaking of chiral symmetry allows $\text{CS}_1(t)$ to vary (see the Supplemental Material \cite{SI} for details).

	\emph{Topological characterizations out of equilibrium.---} 
	We have determined general features of the dynamics of the bulk index. To what extent does this bulk index encode {\it topological} features of the time-evolving state? 
	One may na{\"i}vely expect that the topology of the state is preserved as long as $\text{CS}_1(t)$ does not vary in time (as occurs for all non-trivial classes other than AIII). However, this approach overlooks the gauge dependence of $\text{CS}_1(t)$. An individual measurement of $\text{CS}_1(t)$ at some time $t$ is still only defined modulo 1. Unlike in equilibrium, this ambiguity cannot be resolved by a symmetry-related gauge choice, since TRS and chiral symmetries are  broken by the dynamics. Therefore, wavefunctions with the same $\text{CS}_1$ modulo 1 cannot be distinguished by the bulk index and are thus topologically equivalent. 
	
	Once we restrict ourselves to consider only $\text{CS}_1(t) \mod 1$, we can determine a new classification of states which can be topologically distinguished out of equilibrium; this is given in the last column of Table \ref{tabClassifciation}. Note that systems in classes DIII and CII must be initialized with $\text{CS}_1 \mod 1 = 0$, and hence all such systems are topologically trivial for $t > 0$. A striking consequence of this is that two initial equilibrium states with different topology can time-evolve into the same wavefunction, even though $\text{CS}_1 \mod 1$ does not exhibit any time-dependence; see the Supplemental Material \cite{SI} for an example in class DIII.
	
	One of the clearest signatures of topological non-triviality in equilibrium is the presence of gapless edge excitations \cite{Essin2011}, connected to the non-trivial bulk index through the bulk-boundary correspondence. These edge modes also manifest themselves within the ground state entanglement spectrum \cite{Li2008}, which mimics any physical edge modes that would be present at a boundary \cite{Fidkowski2010, Qi2012}. In the present non-equilibrium setting, the many-body wavefunction $\ket{\Psi(t)}$ can be thought of as the ground state of some fictitious Hamiltonian $\hat{\mathcal{H}}^\text{fic}(t)$ which possesses the same symmetries as the state. For concreteness we can choose (in a second-quantized language) \cite{Vidmar2017,Liu2018}
	\begin{align}
	\hat{\mathcal{H}}^\text{fic}(t) = \hat{\mathcal{U}}(t)\, \hat{\mathcal{H}}^{\text{i}}\, \hat{\mathcal{U}}(t)^\dagger,
	\label{eqFicHamiltonian}
	\end{align}
	\noindent where $\hat{\mathcal{U}}(t)$ is the many-body time evolution operator. The equilibrium entanglement spectrum is a property of the ground state only; therefore the entanglement spectrum of $\ket{\Psi(t)}$ encodes the equilibrium topology of $\hat{\mathcal{H}}^\text{fic}(t)$, which is independent of our specific choice \eqref{eqFicHamiltonian}. If $\hat{\mathcal{H}}^\text{fic}(t)$ cannot be deformed to some trivial Hamiltonian without breaking the enforced symmetries, then it must possess gapless boundary modes \cite{Essin2011}, which themselves will show up in the entanglement spectrum of $\ket{\Psi(t)}$ -- this allows us to probe the bulk-boundary correspondence out of equilibrium.
	
	We now apply the equilibrium classification to $\hat{\mathcal{H}}^\text{fic}(t)$, which due to dynamically-induced symmetry breaking will at most possess PHS only. When PHS is enforced, $\hat{\mathcal{H}}^\text{fic}(t)$ will be topological if and only if the CS invariant of its ground state $\ket{\Psi(t)}$ is a half-odd-integer. We conclude that in 1D a vanishing entanglement gap $\Delta_E$ may only be supported for $t > 0$ in PHS systems which are initialized with a non-integer CS invariant. This is exactly the condition for topological non-triviality that we deduced purely from $\text{CS}_1(t)$, summarized in the last column of Table \ref{tabClassifciation}. 
	Thus we expect the bulk-boundary correspondence to hold out of equilibrium, once CS$_1(t)$ is interpreted modulo 1.

	We have verified these predictions by numerical calculations of the time-evolution of the entanglement spectrum for all symmetry classes in 1D. Results for the contrasting cases of classes AIII and BDI are shown in Fig.\@ \ref{figCurrent}(b). Whilst our arguments have focused on translationally invariant non-interacting systems, our results on the entanglement spectrum and topological classification should be robust against symmetry-preserving disorder, as well as weak interactions.

	In passing, we note that the quench protocol we have used throughout includes Floquet systems as a subset. Indeed, in that context PHS is found to play a different role to TRS and chiral symmetries \cite{Kitagawa2010,Roy2017}. Our results show that the connection between bulk indices and particle transport, which appears in Floquet systems as adiabatic pumping \cite{Kitagawa2010}, holds much more generally, not requiring periodicity or adiabaticity. However, our topological characterization of the instantaneous wavefunction is distinct from the recently classified Floquet SPT orders \cite{Else2016,vonKeyserlingk2016,Potter2016}, which refer to micromotion over a whole period, and cannot be inferred from, e.g.\@ the entanglement spectrum at some fixed time \cite{Potter2016}.
	The preservation of entanglement degeneracies in class D Floquet systems (where no dynamically-induced symmetry breaking occurs) has also previously been observed numerically \cite{Yates2017}.

	\emph{Interacting SPT phases.---} Our consideration of non-interacting fermionic phases reveals the existence of a non-equilibrium topological classification which differs from equilibrium. One expects a similar non-equilibrium classification also for interacting systems, e.g. SPT phases of bosons protected by more general symmetries. Indeed dynamically-induced symmetry breaking, which is a crucial ingredient, can occur in any system: we show in the Supplemental Material \cite{SI} that if a symmetry of the Hamiltonians is realised by an \emph{antiunitary} second-quantized operator $\hat{\mathcal{O}}$ \footnote{Whilst the first-quantized PHS and chiral symmetries in \eqref{eqFirstQuantizedSymmetries} anticommute with the matrix $\matr{H}$, they can be represented as many-body operators which commute with the operator $\hat{\mathcal{H}} = \hat{\psi}_i^\dagger H_{ij} \hat{\psi}_j$, and are unitary and antiunitary on the many-body Hilbert space, respectively. See section II of Ref.~\onlinecite{Chiu2016} for a more complete discussion.}, then $\ket{\Psi(t)}$ will generically not respect that symmetry. Of the three symmetries considered in the main text, only PHS is unitary \cite{Ludwig2016}, and so the results agree.
	
	Unlike for free fermions, a universal bulk index analogous to \eqref{eqChernSimons} does not exist for all 1D SPT phases. Nevertheless, one can still derive a non-equilibrium classification of SPT phases (using e.g.\@ projective symmetry representations \cite{Turner2011,Chen2011,Chen2011b}) which will be reflected in the dynamics of the entanglement spectrum. For example, in the case where only one symmetry is present, it is clear that topology is lost (preserved) if the symmetry is antiunitary (unitary).
	
	%     since the fictitious Hamiltonian \eqref{eqFicHamiltonian} is generally applicable}, the symmetries of $\ket{\Psi(t)}$ are dictated by the symmetries of $\hat{\mathcal{H}}^\text{fic}(t)$.
	
	% Given a symmetry operator $\hat{\mathcal{O}}$ that commutes  with $\hat{\mathcal{H}}^\text{i}$ and $\hat{\mathcal{H}}^\text{f}(t)$, the fictitious Hamiltonian will obey
	% 	\begin{align}
	% 	\hat{\mathcal{O}} \hat{\mathcal{H}}^\text{fic}(t) \hat{\mathcal{O}}^{-1} = \hat{\mathcal{H}}^\text{fic}(\pm t)
	% 	\label{eqAntiuSym}
	% 	\end{align}
	% 	\noindent where the $+$($-$) sign corresponds to the case where $\hat{\mathcal{O}}$ acts as a unitary (anti-unitary) operator on the many-body Fock space. The above can be verified using the property $\hat{\mathcal{O}} i \hat{\mathcal{O}} = \pm i$, but we provide a more rigorous derivation in the Supplemental Material \cite{SI}. Therefore $\hat{\mathcal{H}}^\text{fic}(t)$ and its ground state $\ket{\Psi(t)}$ lose their symmetries when the second-quantized symmetry operator is antiunitary. Of the three symmetries considered in the main text, only PHS is unitary \cite{Ludwig2016}, and so the results agree.
	
	We demonstrate this behavior for the spin-1 Haldane phase, which can be protected by a unitary dihedral symmetry, or by antiunitary TRS \cite{Pollmann2010}. We have numerically investigated the fate of entanglement degeneracies after a quench that does not explicitly break symmetry. We plot the results in Fig.\@ \ref{figCurrent}(c).  (Details are given in the Supplemental Material \cite{SI}.) In the TRS-protected case, the entanglement degeneracy is lifted for $t > 0$, indicating the expected breakdown of the Haldane phase due to dynamically-induced symmetry breaking.
	
	%\MM{Although a universal bulk index such as \eqref{eqChernSimons} does not exist for all 1D SPT phases, we can classify non-equilibrium topology for interacting phases using projective symmetry representations \cite{Turner2011,Chen2011,Chen2011b}. We expect that phases for which an antiunitary symmetry is realised projectively will be unstable after time evolution.}
	
	% 	Since the destruction of TRS after time evolution (along with other antiunitary symmetries) is a generic phenomenon, we expect that TRS-protected SPT phases will also be unstable against unitary dynamics. As a test of this hypothesis, we have numerically investigated the fate of entanglement degeneracies in the spin-1 Haldane phase, which can be protected by any one of several symmetries \cite{Pollmann2010}. 	
	% We consider one case where the phase is protected by a unitary dihedral symmetry, and another case where it is protected only by the antiunitary TRS. The entanglement gap dynamics are plotted in Fig.\@ \ref{figCurrent}(c). In the TRS-protected case, the entanglement degeneracy is lifted for $t > 0$, indicating the breakdown of the Haldane phase due to dynamically-induced symmetry breaking.
	
	In summary we have studied the role played by symmetries in the topological classification of 1D systems that are out of equilibrium, and identified the important phenomenon of dynamically-induced symmetry breaking. It will be of interest to extend these studies to other dimensions and to classify all SPT phases out of equilibrium in future work.
	
	\acknowledgements{We gratefully acknowledge helpful discussions with Jan Carl Budich and Claudio Chamon. This work was supported by an EPSRC studentship and EP/P009565/1, and by the Simons Foundation.  Statement of compliance with EPSRC policy framework on research data: All data are directly available within the publication.}

	\bibliography{topology_ooe_final}
	
	\newpage
	
	%\clearpage
	
	%\pagenumbering{gobble}
	
	\appendix
	
	\setcounter{figure}{0}
	\makeatletter 
	\renewcommand{\thefigure}{S\arabic{figure}}
	
	\newcounter{defcounter}
	\setcounter{defcounter}{0}
	
	\newenvironment{myequation}
	{%
		\addtocounter{equation}{-1}
		\refstepcounter{defcounter}
		\renewcommand\theequation{S\thedefcounter}
		\align
	}
	{%
		\endalign
	}
	
	\begin{onecolumngrid}
		\begin{center}
			{\fontsize{12}{12}\selectfont
				\textbf{Supplemental Material for ``\papertitle''\\[5mm]}}
			{\normalsize Max McGinley and Nigel R. Cooper\\[1mm]}
			{\fontsize{9}{9}\selectfont  
				\textit{\tcm}}
		\end{center}
		\normalsize
	\end{onecolumngrid}
	
	\vspace{20pt}
	
	\begin{twocolumngrid}
		
		\section{Gauge dependence of the CS invariant out of equilibrium}
		
		In the main text, we showed that the fractional part of the CS invariant remains fixed to zero for all time in classes CII and DIII, and then argued that the integer-valued part no longer has physical significance due to the breaking of TRS and chiral symmetries. Here we elaborate on these arguments.
		
		When discussing the notion of topology for non-equilibrium states, we wish to refer \textit{only} to properties of the instantaneous wavefunction $\ket{\Psi(t)}$. We note that topological markers can be identified which characterize the entire history of the wavefunction from time 0 to $t$ \cite{Budich2016}, but our interest is in the former class of invariants. Indeed if one were to make reference to the state at all times, one could construct time-dependent symmetry operators such as $\tilde{\matr{T}}(t) = \matr{U}(t) \matr{T} \matr{U}(t)^T$ as introduced in Ref.\@ \cite{Liu2018}. These `auxiliary' symmetries require knowledge of the entire history of the Hamiltonian $\matr{H}^\text{f}(t)$, and are satisfied only at one instant in time $t$.
		
		If such an operator is defined, we can use it in place of the original TRS operator to construct a gauge in which the parity of the CS invariant is a topological invariant. However, such an invariant is a characterization of the full-time trajectory of the state, since $\tilde{\matr{T}}(t)$ can only be defined if $\matr{H}^\text{f}(t)$ is known. More physically, if one were to construct this invariant in practice, one could measure the equilibrium topological invariant at $t=0$ and then monitor the fractional part of the CS invariant continuously in time. If we choose the integer part of $\text{CS}_1(t)$ such that its full value has no discontinuities in time, we will arrive at an unambiguously defined invariant which is equivalent to the invariant constructed using $\tilde{\matr{T}}(t)$. It is clear that information beyond that of the instantaneous $\ket{\Psi(t)}$ must be known in order to construct such an invariant.
		
		The above has a rather striking consequence in that two topologically distinct states in class DIII at $t=0$ can both reach the \textit{same} state after time evolution, even when $\matr{H}^\text{f}(t)$ respects all the required symmetries. We provide an example of such a scenario, using a spin-half Bogoliubov-de Gennes model in 1D. Denoting the Pauli operators in spin space as $\sigma^\alpha$ for $\alpha = \{x,y,z\}$ and in Nambu space as $\tau^\alpha$, we ensure that all Hamiltonians respect TRS with $\matr{T} = i\sigma^y\otimes 1$; PHS with $\matr{C} = 1\otimes \tau^x$; and chiral symmetry with $\matr{S} = \sigma^y \otimes \tau^x$. Now consider two quench protocols $A$ and $B$. In protocol $A$ we start in a topologically trivial state, and then quench to a new time-independent final Hamiltonian
		
		\begin{myequation}
			\matr{H}^\text{i}_A(k) = 1 \otimes\tau^z \;\; \rightarrow \;\;\matr{H}^\text{f}_A(k,t) = \sigma^y\otimes \tau^y 
		\end{myequation} 
		
		In protocol $B$ we start in a topologically non-trivial state (which is equivalent two time-reversal copies of a topological $p$-wave wire \cite{Kitaev2001}), and then quench to another time-independent Hamiltonian
		\begin{myequation}
			\matr{H}^\text{i}_B(k) &= 1\otimes\left[ \cos (k) \tau^z + \sin(k) \tau^y\right] \nonumber\\ \rightarrow \;\; \matr{H}^\text{f}_B(k,t) &= \sigma^y\otimes \left[-\sin (k) \tau^z + \cos (k) \tau^y \right]
		\end{myequation}
		
		Because both initial states are independent of spin indices, the time-evolved states are also spin-independent, and can be described by $k$-dependent spinors in Nambu space which have precessed under a $k$-dependent field defined by the final Hamiltonian. At time $t=\pi/2$, the states of system $A$ and $B$ will be the same, corresponding to all spinors pointing along $\tau^x$. Even though both systems started in topologically distinct initial states and evolved under class DIII Hamiltonians, the two systems have been brought to the same final state. Furthermore, if one were to compute the invariant using the auxiliary symmetry $\tilde{\matr{T}}(t)$ described above, one would find that system $A$ has an even CS invariant whilst system $B$ has an odd CS \emph{even though their final states are the same}. Indeed the different initial states affect the full-time characterization of the quench.
		
		We also note that the time-dependent auxiliary symmetries $\tilde{\matr{T}}(t)$ are in general $k$-dependent for $t >0$, i.e.  they have non-local spatial profile. Therefore when we consider the entanglement spectrum,
		we need not consider the effect of these symmetries, since an entanglement cut will not respect a symmetry which is not a product of on-site unitary operators.
		
		\section{Explicit symmetry breaking in the SSH model}
		
		The presence of TRS, PHS, and chiral symmetries is often referred to as a binary question: a system either does or does not possess each of the symmetries. However, whilst there exist canonical forms for each of the symmetry operators, (e.g. $\matr{T} = 1$ in a spinless system), it is possible for two systems to realize the same symmetry with different unitary matrices $\matr{T}, \matr{C}, \matr{S}$, as defined in Eq.\@ \eqref{eqFirstQuantizedSymmetries}. A simple yet rather subtle case arises in the SSH model in which TRS can be realized in more than one form.
		
		The SSH Hamiltonian is \cite{Su1979}
		
		\begin{myequation}
			H = -\sum_j J_1\hat{c}_{2j+1}^\dagger \hat{c}_{2j} + J_2\hat{c}_{2j+2}^\dagger \hat{c}_{2j+1} + \text{h.c.},
		\end{myequation}
		
		\noindent which is equivalent to the model defined in the main text with $B_1 = B_2 = 0$. If we use a 2-site unit cell $\{2j, 2j+1\}$, the Hamiltonian has a momentum-space representation
		
		\begin{myequation}
			\matr{H}(k) = \begin{pmatrix}
				0 & J_1 + J_2^* e^{-ik} \\
				J_1^* + J_2 e^{ik} & 0
			\end{pmatrix}.
		\end{myequation}
		
		When the hopping amplitudes $J_{1,2}$ are real, then the system respects TRS with $\matr{T} = 1$, since $\matr{H}(k)^* = \matr{H}(-k)$. Clearly if the hopping amplitudes become complex then this is no longer true. It has previously been stated \cite{Velasco2017} that the SSH model with complex hopping amplitudes does not respect any TRS and is thus class AIII, since there is no $2\times 2$ matrix $\matr{T}$ which satisfies $\matr{T} \matr{H}(k)^* \matr{T}^\dagger = \matr{H}(-k)$. However, such a condition is too stringent: a 2-site unit cell has been implicitly assumed. The TRS condition should not rely on the existence of a unit cell at all, since disordered systems can still be time-reversal symmetric.
		
		If we instead look for a unitary matrix $\matr{T}$ in real space which is a TRS in the sense of Eq.\@ \eqref{eqFirstQuantizedSymmetries}, we could define
		
		\begin{myequation}
			\matr{T} = \text{diag}\left(\cdots, 1, e^{-2i\phi_1}, e^{-2i(\phi_1+\phi_2)}, e^{-2i(2\phi_1 + \phi_2)}, \right.\nonumber\\ \left.  e^{-2i(2\phi_1 + 2\phi_2)}, e^{-2i(3\phi_1 + 2\phi_2)}, \cdots \right)
			\label{eqTRSSSH}
		\end{myequation}
		
		\noindent where $\phi_{1,2}$ are the phases of the complex hoppings $J_{1,2}$. This is perhaps not surprising given that the Hamiltonian is equivalent to the real-amplitude model through the gauge transform $c_{2j} \rightarrow c_{2j}e^{-i(\phi_1+\phi_2)j}; c_{2j+1} \rightarrow c_{2j+1}e^{-i(\phi_1+\phi_2)j}e^{-i\phi_1}$. In a similar fashion to the Hofstadter model \cite{Hofstadter1976}, the presence of the hopping phases forces us to extend the unit cell from the na{\"i}ve choice, which is why the $2 \times 2$ matrix approach fails to detect this TRS. When the phase $(\phi_1 + \phi_2)$ is an integer multiple of $\pi$, the diagonal entries of $\matr{T}$ have a periodicity of 2 sites; however, if $(\phi_1 + \phi_2)/\pi = p/q$ with $p$ and $q$ coprime integers, then we must define a unit cell with $2q$ sites. Regardless, a TRS operator always exists, and so one may call the SSH model with complex hopping amplitudes a class BDI system. A PHS $\matr{C} = \matr{S}\matr{T}$ will also be defined in this model.
		
		The TRS operator \eqref{eqTRSSSH} and the corresponding PHS are highly non-generic symmetries which are specific to the choice of phase of both hopping amplitudes. If we consider a quench from one complex SSH model to another, then if the phases of either $J_{1,2}$ change across the quench, the TRS and PHS will be explicitly broken since the initial and final Hamiltonians realize different symmetries. The same chiral (sublattice) symmetry is present in the Hamiltonian at all times but will be broken by the dynamics, yielding class AIII-like behaviour with a time-varying bulk index. On the other hand, if the phases $\phi_{1,2}$ remain the same across the quench then the TRS and PHS are not explicitly broken: one will see class BDI behaviour with a constant-in-time CS invariant.
		
		\section{Symmetries of the time-evolved wavefunction}
		
		Here we demonstrate that any symmetry of the Hamiltonian which is realised by an antiunitary operator will not be reflected in the time-evolved state. Heuristically, when $\hat{\mathcal{H}}(t)$ is time-independent one can see this easily since the factor of $i$ in the time evolution operator $\hat{\mathcal{U}}(t)$ is not invariant under an antiunitary operator, but we prove this rigorously in a way which is also valid for time-dependent $\hat{\mathcal{H}}(t)$.
		
		We start with the time-dependent Schr{\"o}dinger equation
		\begin{myequation}
			\frac{\partial}{\partial t} |\Psi(t)\rangle = -i\hat{\mathcal{H}}(t) |\Psi(t)\rangle,
			\label{eqTDSE}
		\end{myequation}
		\noindent where the many-body Hamiltonian $\hat{\mathcal{H}}(t)$ commutes with a symmetry operator $\hat{\mathcal{O}}$ that is unitary or antiunitary. By definition, $\hat{\mathcal{O}}$ satisfies $\braket{\hat{\mathcal{O}}\Phi, \hat{\mathcal{O}}\Psi} = \braket{\Phi,\Psi}^{(*)}$ for any two states $\ket{\Phi}$ and $\ket{\Psi}$, where $\braket{\cdot, \cdot}$ is the inner product on the Hilbert space.  Here, $^{(*)}$ denotes complex conjugation in the antiunitary case, and is absent in the unitary case. The condition for $\ket{\Phi}$ to start in an `$\hat{\mathcal{O}}$-symmetric' state is that $\hat{\mathcal{O}}\ket{\Psi(t)} = \eta \ket{\Psi(t)}$, where $\eta$ is real.
		
		Now, taking the inner product of \eqref{eqTDSE} with an arbitrary state $\ket{\Phi}$ and using linearity of the inner product in the second argument gives
		\begin{myequation}
			\frac{\partial}{\partial t} \braket{\Phi, \Psi(t)} &= -i\braket{\Phi, \hat{\mathcal{H}}(t) \Psi(t)}
		\end{myequation}
		\noindent which we wish to compare to the time evolution of $\hat{\mathcal{O}} |\Psi(t)\rangle$. We can compare the above to the equivalent expression for the state $\hat{\mathcal{O}} |\Psi(t)\rangle$, which gives
		\begin{myequation}
			\frac{\partial}{\partial t}\braket{\Phi, \hat{\mathcal{O}} \Psi(t)} &= \frac{\partial}{\partial t} \left\langle \hat{\mathcal{O}}^{-1}\Phi, \Psi(t)\right\rangle ^{(*)} \nonumber\\ &= \left\langle \hat{\mathcal{O}}^{-1}\Phi, \frac{\partial}{\partial t} \Psi(t)\right\rangle ^{(*)} \nonumber\\
			&= \left\langle \hat{\mathcal{O}}^{-1}\Phi, -i\hat{\mathcal{H}}(t) \Psi(t)\right\rangle ^{(*)} \nonumber\\
			&= \mp i \left\langle \hat{\mathcal{O}}^{-1}\Phi, \hat{\mathcal{H}}(t) \Psi(t)\right\rangle ^{(*)} \nonumber\\
			&= \mp i \left\langle \Phi,\, \hat{\mathcal{O}} \hat{\mathcal{H}}(t) \Psi(t)\right\rangle  \nonumber\\
			&= \mp i \left\langle \Phi,\, \hat{\mathcal{H}}(t) \hat{\mathcal{O}}  \Psi(t)\right\rangle.
		\end{myequation}
		The first step follows from the (anti-)unitarity of $\hat{\mathcal{O}}^{-1}$. We use (anti-)linearity in the second argument of $\langle \cdot, \cdot \rangle^{(*)}$ going from the third to fourth line, and $[\hat{\mathcal{O}}, \hat{\mathcal{H}}(t)] = 0$ in the last step.  Because this is true for arbitrary $\ket{\Phi}$, we therefore conclude that
		\begin{myequation}
			\frac{\partial}{\partial t} \left[\hat{\mathcal{O}} |\Psi(t)\rangle\right] = \mp i \hat{\mathcal{H}}(t) \left[\hat{\mathcal{O}}  |\Psi(t)\rangle\right],
		\end{myequation}
		Comparing the above with \eqref{eqTDSE}, we see that the state $\hat{\mathcal{O}} \ket{\Psi(t)}$ evolves under the Hamiltonian $\pm \hat{\mathcal{H}}(t)$. Thus when $\hat{\mathcal{O}}$ is antiunitary, $\hat{\mathcal{O}} \ket{\Psi(t)} \neq \eta \ket{\Psi(t)}$ for any $\eta$ and thus the symmetry is generically lost.
		
		\section{Details of numerical calculations of the Haldane phase}
		
		The results presented in Figure \ref{figCurrent}(c) were calculated with the help of the iTensor package \cite{ITensor}. We use the DMRG algorithm to find the ground state of the initial Hamiltonian, and TEBD to time-evolve this state under the final Hamiltonian. Throughout, we use a system size of $N = 48$ with open boundary conditions and retain a bond dimension of $\chi = 100$, which allows us to reach a time $t \approx 3J_1$ before appreciable truncation errors set in.
		
		The Hamiltonian which we use in our simulation is based on the AKLT Hamiltonian, with various symmetry-violating terms added:
		
		\begin{myequation}
			\hat{\mathcal{H}} &= \hat{\mathcal{H}}_\text{AKLT} + \hat{\mathcal{H}}_{zz} + \hat{\mathcal{H}}_\text{inv} + \hat{\mathcal{H}}_\text{di}; \nonumber\\
			\hat{\mathcal{H}}_\text{AKLT} &= J_1 \sum_j \vec{S}_j \cdot \vec{S}_{j+1} + \frac{1}{3}(\vec{S}_j \cdot \vec{S}_{j+1})^2, \nonumber\\
			\hat{\mathcal{H}}_{zz} &= U_{zz} \sum_j (S_j^z\, S_{j+1}^z)^2, \nonumber\\
			\hat{\mathcal{H}}_\text{inv} &= R\sum_j (S^x_j S^y_{j+1})^2 - (S^y_j S^x_{j+1})^2, \nonumber\\
			\hat{\mathcal{H}}_\text{di} &= D \sum_j S^x_j S^z_{j+1} - S^z_j S^x_{j+1}.
		\end{myequation}
		
		The AKLT Hamiltonian belongs to the Haldane phase, and respects inversion symmetry $\vec{S}_j \rightarrow \vec{S}_{-j+1}$, dihedral symmetry of $\pi$ rotations about each of the Cartesian axes, and time reversal symmetry $\vec{S}_j \rightarrow -\vec{S}_j$. Any one of these three symmetries is enough to protect the Haldane phase. The term $\hat{\mathcal{H}}_{zz}$ is added to provide a quench parameter; $\hat{\mathcal{H}}_\text{inv}$ breaks inversion symmetry, and $\hat{\mathcal{H}}_\text{di}$ breaks both inversion and dihedral symmetry. Even with all parameters non-zero, the system still possesses a Haldane phase protected by TRS.
		
		The green line in Figure \ref{figCurrent}(c) corresponds to $D=0$, possessing the unitary dihedral symmetry, and the quench chosen is $(J_1, U_{zz}, R) = (1,0.2,0.15) \rightarrow (0.5, 1.7, 0.15)$. The purple line follows the same quench with the exception that $D = 0.15$ throughout, thus breaking the dihedral symmetry. When the Haldane phase is protected by TRS alone, the entanglement degeneracy is lifted for $t > 0$.
		
	\end{twocolumngrid}
	
\end{document}